\documentclass[rnote,traditabstract]{aa} 
\usepackage{graphicx}
\usepackage{natbib}
\usepackage{lscape}
\usepackage{longtable}
\usepackage{multirow}
\usepackage{url}

\usepackage{txfonts}
\begin{document}
\newcommand{\Zsolar}{\mbox{$\,\rm Z_{\odot}$}}
\newcommand{\Msolar}{\mbox{$\,\rm M_{\odot}$}}
\newcommand{\Lsolar}{\mbox{$\,\rm L_{\odot}$}}
\newcommand{\xs}{$\chi^{2}$}
\newcommand{\dxs}{$\Delta\chi^{2}$}
\newcommand{\xsn}{$\chi^{2}_{\nu}$}
\newcommand{\ls}{{\tiny \( \stackrel{<}{\sim}\)}}
\newcommand{\gs}{{\tiny \( \stackrel{>}{\sim}\)}}
\newcommand{\asec}{$^{\prime\prime}$}
\newcommand{\amin}{$^{\prime}$}
\newcommand{\mstar}{\mbox{$M_{*}$}}
\newcommand{\hi}{H{\sc i}\ }
\newcommand{\hii}{H{\sc ii}\ }
\newcommand{\kms}{$\rm km~s^{-1}$}

   \title{Are passive red spirals truly passive?}
   \subtitle{The current star formation activity of optically-red disc galaxies}
\author{L. Cortese}

\institute{
European Southern Observatory, Karl Schwarzschild Str. 2, 85748 Garching bei M\"unchen, Germany
}

\date{Received 19 April 2012 - Accepted 30 May 2012}

 
  \abstract{We use GALEX ultraviolet and WISE 22 $\mu$m observations to investigate the current star formation activity  of the 
  optically-red spirals recently identified as part of the Galaxy Zoo project. These galaxies were accurately selected from the Sloan Digital 
  Sky Survey in order to be pure discs with low or no current star formation activity, representing one of the best 
  optically-selected samples of candidate passive spirals. However, we show that these galaxies are not only still forming stars 
  at a {\it significant} rate ($\ga$ 1 M$_{\odot}$ yr$^{-1}$) but, more importantly, their star formation activity is not different from that of 
  normal star-forming discs of the same stellar mass (M$_{*}\ga$ 10$^{10.2}$ M$_{\odot}$). 
  Indeed, these systems lie on the UV-optical blue sequence, even without any corrections for internal dust attenuation, and 
  they follow the same specific star formation rate vs. stellar mass relation of star-forming galaxies. 
  Our findings clearly show that, at high stellar masses, optical colours do not allow to discriminate between 
  actively star-forming and truly quiescent systems.}


	\authorrunning{Cortese}	
	
   \maketitle
%

\section{Introduction}
Since its discovery, the colour vs. magnitude (or stellar mass) relation 
(e.g., \citealp{visvanathan77,visvanathan81}) has been adopted as one of the 
most useful tools to characterize the properties of galaxies across the 
Hubble time. Particularly powerful is the use of colours to discriminate between 
late-type/star-forming systems and early-type/passive galaxies, when accurate morphological 
classification is not available. However, it is now well known that the bimodality in the colour 
distribution of galaxies does not always reflect a difference in morphological type \citep{scodeggio02,franzetti07}. 
Indeed, at least in the local Universe, blue/star-forming spheroids (e.g., \citealp{yi05,kannappan09,schawinski09}) and 
red/quiescent spirals (e.g., \citealp{vandberg76,moran06,crowl2008,cortese09,wolf09,bundy10,rowlands12}) do exist.

In the last decade, we have seen a rapid increase in the number of studies 
focused on such unusual systems with the aim to understand why they do not follow the typical 
relations between morphology and colours. 
However, while identifying current star formation activity in elliptical galaxies is 
relatively easy, isolating truly quiescent disc galaxies is a more complicated issue. 
Indeed, optical red colours do not always automatically imply a passive stellar 
population (e.g., \citealp{cowie08,cortese09}). 
Dust extinction and a large bulge component are among some of the issues that 
can significantly affect the observed optical colours of discs, making them appear 
much more `evolved' then they really are. For these reasons, the vast majority of the works focused 
on truly passive spirals have taken advantage of multiwavelength datasets spanning from the ultraviolet 
to the mid- and far-infrared regime, in order to account at least for the effects 
of internal dust absorption (e.g., \citealp{wolf09,gallazzi09}).

Recently, as part of the Galaxy Zoo project \citep{lintott08}, \citet[hereafer M10]{masters10} 
performed a careful selection to isolate and study red spirals. 
They take advantage of the morphological classification available for 
thousands of objects to select spiral galaxies and then adopt 
a simple cut in inclination, optical colour and bulge fraction to 
identify disc-dominated objects which should be `truly passive' (i.e. 
with current star formation activity significantly lower than the one 
observed in normal spiral galaxies of similar stellar mass). 
The novelty of this technique lies in the fact that only optical data, 
combined with an accurate morphological classification, are needed 
to isolate the red spiral population. 
If successful, these selection criteria may make significantly easier 
to isolate passive systems, thus improving our understanding of the evolutionary 
paths leading to such unusual galaxies. 

Thus, in this Research Note we take advantage of Galaxy Evolution Explorer (GALEX, \citealp{martin05}) 
ultraviolet (UV) and Wide-field Infrared Survey Explorer (WISE, \citealp{wright10}) 22 $\mu$m observations to study the star formation properties of these 
optically-red spiral galaxies and to determine whether they are really passive.


\section{The data}
The sample of optically-red spirals presented by M10 comes from the 
Galaxy Zoo clean catalogue \citep{lintott08}, which is based on the Sloan Digital Sky Survey (SDSS) DR6 \citep{sdss_DR6}. 
First, a volume-limited sample of spirals was obtained assuming 
spiral likelihood\footnote{This corresponds to the fractional number of cases 
in which a galaxy has been classified as a spiral by the public.} \citep{bamford09} greater than 0.8, spectroscopic redshift 0.03$<z<$0.085 and $r$-band 
absolute magnitude $M_{r}<-$20.17 mag. Secondly, bulge-less and face-on discs were selected 
requiring a ratio between the major and minor axis of the galaxy $\log(a/b)<$0.2 and a fraction 
of the light coming from the deVaucoleur component of the surface brightness profile $fracDev<$0.5. 
Finally, red spirals were defined as those objects having a color $(g-r)>$0.63-0.02$\times$($M_{r}$+20). 
The sample of red spirals so obtained includes 294 galaxies. 

Unfortunately, not all the galaxies in the M10 sample are included in the GALEX GR6 public release. 
Thus, in the following analysis, we only focus on the subset of 255 galaxies observed by GALEX and lying within 
0.55\degr from the center of a GALEX tile, in order to avoid edge effects\footnote{Only 7 galaxies in the sample have been observed 
but lie at distances greater than 0.55\degr.}. 
Given that this subset includes $\sim$87\% of the original sample, we are confident that our conclusions 
can be extended to the whole population of optically-red galaxies studied by M10.

\subsection{GALEX \& WISE photometry}
We cross-correlated the 255 red spirals with the WISE all-sky survey source catalogue using a matching radius of 10 arcsec. We found 233 galaxies detected at 22 $\mu$m but, 
after a careful visual inspection of the images, 46 turned out to be spurious or marginal detections, or in regions 
highly contaminated by foreground emission. 
Thus, out of the 255 galaxies in our sample, 166 ($\sim$65\%) are clearly detected at 22 $\mu$m.

In order to obtain homogeneous flux density estimates in the GALEX UV and WISE 22 $\mu$m bands, we performed aperture 
photometry using the same aperture size in all images. We adopted circular apertures of size twice the isophotal diameter 
at the 25 mag arcsec$^{-2}$ level in $r$ band, convolved to the WISE 22 $\mu$m resolution (12 arcsec). In a few cases, the size of the aperture has been 
adjusted to make sure that all the emission from the galaxy is included. Sky background was determined in circular 
annuli or boxes around the target, and foreground/background objects were accurately masked. 

All the 255 galaxies in our sample are detected in near-ultraviolet (NUV), while only for 220 objects 
(86\% of the sample) we were able to obtain a far-ultraviolet (FUV) flux density. This is mainly because the FUV observations are 
in general less sensitive than the NUV ones.   
By comparing the magnitudes obtained from independent observations of the same galaxy, as well as 
with the GALEX GR6 public catalogue, we obtain a typical uncertainty of $\sim$0.15 and 0.30 mag in NUV and FUV, 
respectively. 
Similarly, we find a very good agreement between our flux densities and the {\it standard} 
aperture estimates (\textsc{w4mag}) in the WISE all-sky survey catalogue, with a typical 
scatter of $\sim$0.2 mag.

UV magnitudes have been corrected for Galactic extinction following \cite{wyder07}.   
No aperture or colour corrections were applied to the 22 $\mu$m flux densities. These are in general 
smaller than the uncertainty on the flux density, and thus do not affect our results.    

\begin{figure}
  \centering
  \includegraphics[width=9cm]{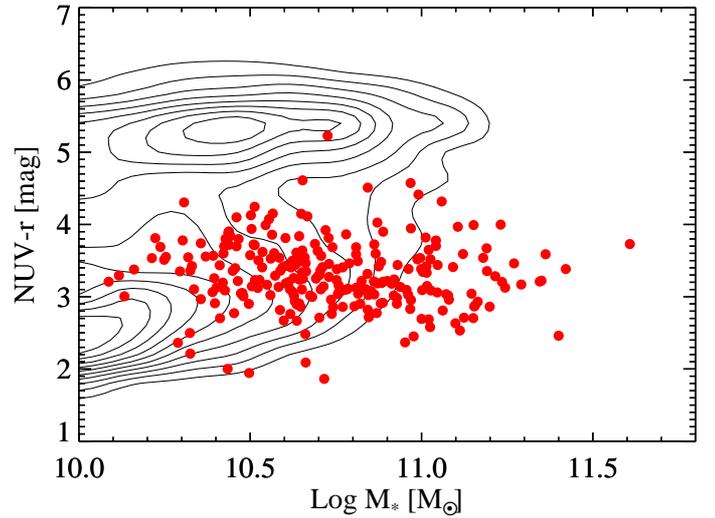}
     \caption{The optically-red spirals (red circles) on a $NUV-r$ vs. stellar mass diagram. The contours show the distribution 
     of galaxies in the GASS survey \citep{catinella10}.}
	 \label{cmd}
  \end{figure}

\section{Results}
Fig.~\ref{cmd} shows where the red spirals (red filled circles) lie in a $NUV-r$ colour vs. stellar mass (M$_{*}$) diagram. 
Stellar masses are taken from the MPA/JHU SDSS DR7 release\footnote{\url{http://www.mpa-garching.mpg.de/SDSS/DR7/}}. These 
are derived from SDSS photometry using the spectral energy distribution fitting technique described in \cite{salim07}, 
assuming a \cite{chabrier} initial mass function (IMF). 
The black contours show the colour distribution of the 
parent sample of the GALEX Arecibo SDSS survey (GASS, \citealp{catinella10}), a volume-limited sample of nearby 
massive galaxies (M$_{*}>$10$^{10}$ M$_{\odot}$), providing an indication of the typical range of UV colours 
of red- and blue-sequence massive galaxies in the local Universe.

It is clear that the red spirals in M10 sample are not quiescent. Not only there is just one galaxy 
in the UV red sequence, but almost all the sample (94\%, i.e., 240 out of the 255 galaxies) have 
an observed $NUV-r$ colour bluer than 4 mag, i.e., the typical value observed in actively star-forming galaxies (e.g., \citealp{salim07}).

In order to determine if the current star formation activity of these optically-red spirals is different from that of 
normal star-forming galaxies of similar stellar mass, we determined star formation rates (SFRs) 
by combining the GALEX UV and WISE 22 $\mu$m data. 
We used the 22 $\mu$m fluxes to correct the NUV and FUV luminosities for dust attenuation following \cite{hao11}:
\begin{equation}
\begin{array}{l}
 L(NUV)_{corr} = L(NUV) + 2.26\times L(22\mu m) \\
 L(FUV)_{corr} = L(FUV) + 3.89\times L(22\mu m) 
\end{array}
\end{equation}
where $L$ are luminosities in units of erg s$^{-1}$. 
It is important to note that these relations were calibrated using 
Spitzer 24 $\mu$m and IRAS 25 $\mu$m flux densities, whereas in this case we are applying them to 
the WISE 22 $\mu$m emission. Given that, for typical star-forming objects, the 22 $\mu$m emission should 
be comparable or slightly fainter than the 24 $\mu$m flux density, we can assume our corrections to be 
a lower limit to the real value. For galaxies observed but not detected by WISE we considered two 
different extreme cases by either fixing the 22 $\mu$m emission to zero or to the typical 3$\sigma$ 
sensitivity limit of WISE (i.e. 3 mJy).    
We then estimated the current SFR from the corrected UV luminosities using the 
relation presented by \cite{salim07} and calibrated on a \cite{chabrier} IMF
\begin{equation}
\label{sfr}
SFR~ (\rm M_{\odot} yr^{-1}) = 1.46 \times 10^{-28} \times  L_{\nu}
\end{equation}
where $L_{\nu}$ is in units of erg s$^{-1}$ Hz$^{-1}$.
The SFRs estimated from the NUV and FUV luminosities agree within $\sim$10-25\%, 
with the NUV-based SFR being typically higher by $\sim$0.03-0.07 dex.   

\begin{figure}
  \centering
  \includegraphics[width=9cm]{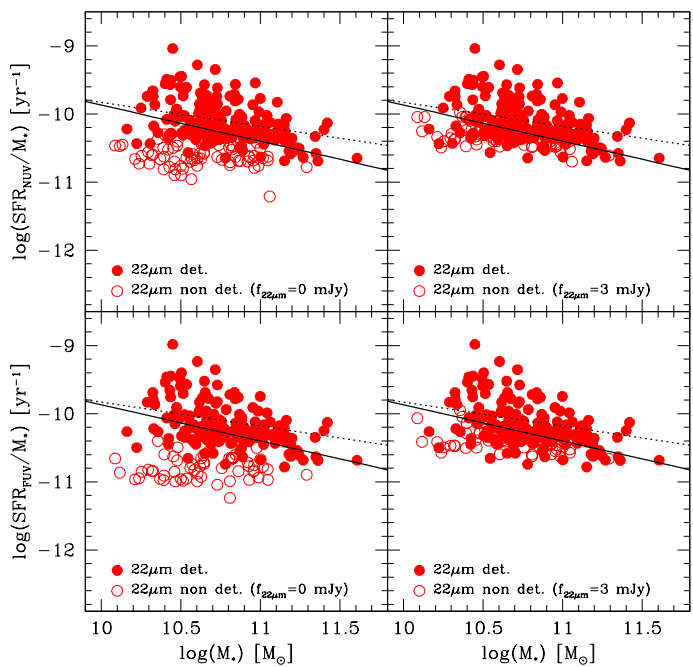}
     \caption{The NUV- (top) and FUV-based (bottom) specific star formation rate vs. stellar mass relation for
     the sample of optically-red spirals.
     Filled and empty circles show 22 $\mu$m detections and non detections, respectively. For non detections, the SFR has been determined by assuming 
     a 22 $\mu$m flux density equal to zero (left panels) or to 3mJy (i.e., $\sim$3$\sigma$ WISE detection limit; right panels).
     The dotted and solid lines show the best-fit to the star-forming sequence obtained by \cite{salim07} by including only star-forming galaxies (i.e., excluding 
     galaxies with optical lines ratios typical of active galactic nuclei) or all galaxies with $NUV-r$ colour lower than 4, respectively.}
	 \label{ssfr}
  \end{figure}
In Fig.~\ref{ssfr} we plot the specific star formation rate ($SSFR=SFR/M_{*}$) as a function of the 
stellar mass for the M10 sample. The top and bottom panels show the results obtained for the NUV-based and FUV-based star formation 
rates, respectively. For each row, the difference between the left and right panels is in the treatment of the WISE non detections. 
While in the left panels the 22 $\mu$m flux density is set to zero, in the right ones it is equal to the typical 3$\sigma$ noise level 
of WISE images. Thus, the two panels show the two extremes, with the real relation followed by this sample lying 
in between the two plots. 

As expected from what already shown in Fig.~\ref{cmd}, and regardless the way we treat non detections, 
all the red spirals are forming stars at a significant rate and $\sim$85-90\% (depending on which 
SFR estimate is used) have a current $SFR\geq$1 M$_{\odot}$ yr$^{-1}$.
More importantly, the M10 sample follows remarkably well the relation obtained 
for nearby star-forming galaxies by \citet[solid and dashed lines in Fig.~\ref{ssfr}]{salim07}, implying that these galaxies 
are not different from the local population of blue sequence discs.

\section{Discussion \& Conclusion}
The results presented in the previous section clearly show that the red spirals in the M10 sample are not passive 
but are still forming stars at the rate expected for their stellar mass. 
It is thus important to briefly discuss why the very careful optical selection performed by M10 did not isolate truly quiescent systems. 

We can safely exclude that dust attenuation plays a dominant role. The selection of face-on spirals guarantees the absence of 
highly obscured systems, as clearly shown in Fig.~\ref{cmd}. Indeed, dust extinction would affect much more heavily the UV colours 
moving all the galaxies to the UV red sequence, contrary to what observed.
 
In order to understand why these objects have optically-red colours, despite their still on-going star formation activity, 
it is important to remember that they are massive galaxies. For stellar masses $\ga$10$^{10}$ M$_{\odot}$, the optical blue 
cloud merges into the red sequence, suggesting that optical colours are no
longer a good proxy for the current star formation 
activity of a galaxy. Indeed, as extensively discussed by \cite{wyder07}, in UV the blue and the red sequences are well separated even at 
high luminosities, implying that massive spirals are forming stars despite what suggested by their optical colours. 
This is, at least partially, a consequence of the fact that massive objects formed the bulk of their stars at earlier epochs 
than dwarf galaxies (e.g., \citealp{boselli01,gav02,heavens,thomas10}), and that optical colours are more directly related to the 
average age of the stellar populations than to the current star formation activity (e.g., \citealp{wyder07,chilingarian12}). 
The fact that the fraction of the optically-red spirals in the M10 sample significantly increases with stellar mass 
(see Fig. 2 of M10) is consistent with this scenario.

In conclusion, at least at high stellar masses, optical colours alone are not sufficient to discriminate between 
actively star-forming and quiescent systems.

\begin{acknowledgements}
I wish to thank Daniel Thomas for encouraging the publication of this study and for useful comments, and Barbara Catinella for providing 
the data from the GASS survey.
 
GALEX (Galaxy Evolution Explorer) is a NASA Small Explorer, launched in April 2003. 
We gratefully acknowledge NASA's support for construction, operation, and science analysis for the GALEX mission, 
developed in cooperation with the Centre National d'Etudes Spatiales (CNES) of France and the 
Korean Ministry of Science and Technology. 

This publication makes use of data products from the Wide-field Infrared Survey Explorer, which is a joint project 
of the University of California, Los Angeles, and the Jet Propulsion Laboratory/California Institute of Technology, 
funded by the National Aeronautics and Space Administration.

The research leading to these results has received funding from the European Community's Seventh Framework Programme 
(/FP7/2007-2013/) under grant agreement No 229517.

\end{acknowledgements}

\bibliography{main}

\begin{thebibliography}{32}
\expandafter\ifx\csname natexlab\endcsname\relax\def\natexlab#1{#1}\fi

\bibitem[{{Adelman-McCarthy} {et~al.}(2008){Adelman-McCarthy}, {Ag{\"u}eros},
  {Allam}, {Allende Prieto}, {Anderson}, {Anderson}, {Annis}, {Bahcall},
  {Bailer-Jones}, {Baldry}, {Barentine}, {Bassett}, {Becker}, {Beers}, {Bell},
  {Berlind}, {Bernardi}, {Blanton}, {Bochanski}, {Boroski}, {Brinchmann},
  {Brinkmann}, {Brunner}, {Budav{\'a}ri}, {Carliles}, {Carr}, {Castander},
  {Cinabro}, {Cool}, {Covey}, {Csabai}, {Cunha}, {Davenport}, {Dilday}, {Doi},
  {Eisenstein}, {Evans}, {Fan}, {Finkbeiner}, {Friedman}, {Frieman},
  {Fukugita}, {G{\"a}nsicke}, {Gates}, {Gillespie}, {Glazebrook}, {Gray},
  {Grebel}, {Gunn}, {Gurbani}, {Hall}, {Harding}, {Harvanek}, {Hawley},
  {Hayes}, {Heckman}, {Hendry}, {Hindsley}, {Hirata}, {Hogan}, {Hogg}, {Hyde},
  {Ichikawa}, {Ivezi{\'c}}, {Jester}, {Johnson}, {Jorgensen}, {Juri{\'c}},
  {Kent}, {Kessler}, {Kleinman}, {Knapp}, {Kron}, {Krzesinski}, {Kuropatkin},
  {Lamb}, {Lampeitl}, {Lebedeva}, {Lee}, {Leger}, {L{\'e}pine}, {Lima}, {Lin},
  {Long}, {Loomis}, {Loveday}, {Lupton}, {Malanushenko}, {Malanushenko},
  {Mandelbaum}, {Margon}, {Marriner}, {Mart{\'{\i}}nez-Delgado}, {Matsubara},
  {McGehee}, {McKay}, {Meiksin}, {Morrison}, {Munn}, {Nakajima}, {Neilsen},
  {Newberg}, {Nichol}, {Nicinski}, {Nieto-Santisteban}, {Nitta}, {Okamura},
  {Owen}, {Oyaizu}, {Padmanabhan}, {Pan}, {Park}, {Peoples}, {Pier}, {Pope},
  {Purger}, {Raddick}, {Re Fiorentin}, {Richards}, {Richmond}, {Riess}, {Rix},
  {Rockosi}, {Sako}, {Schlegel}, {Schneider}, {Schreiber}, {Schwope}, {Seljak},
  {Sesar}, {Sheldon}, {Shimasaku}, {Sivarani}, {Smith}, {Snedden}, {Steinmetz},
  {Strauss}, {SubbaRao}, {Suto}, {Szalay}, {Szapudi}, {Szkody}, {Tegmark},
  {Thakar}, {Tremonti}, {Tucker}, {Uomoto}, {Vanden Berk}, {Vandenberg},
  {Vidrih}, {Vogeley}, {Voges}, {Vogt}, {Wadadekar}, {Weinberg}, {West},
  {White}, {Wilhite}, {Yanny}, {Yocum}, {York}, {Zehavi}, \&
  {Zucker}}]{sdss_DR6}
{Adelman-McCarthy}, J.~K., {Ag{\"u}eros}, M.~A., {Allam}, S.~S., {et~al.} 2008,
  \apjs, 175, 297

\bibitem[{{Bamford} {et~al.}(2009){Bamford}, {Nichol}, {Baldry}, {Land},
  {Lintott}, {Schawinski}, {Slosar}, {Szalay}, {Thomas}, {Torki}, {Andreescu},
  {Edmondson}, {Miller}, {Murray}, {Raddick}, \& {Vandenberg}}]{bamford09}
{Bamford}, S.~P., {Nichol}, R.~C., {Baldry}, I.~K., {et~al.} 2009, \mnras, 393,
  1324

\bibitem[{{Boselli} {et~al.}(2001){Boselli}, {Gavazzi}, {Donas}, \&
  {Scodeggio}}]{boselli01}
{Boselli}, A., {Gavazzi}, G., {Donas}, J., \& {Scodeggio}, M. 2001, \aj, 121,
  753

\bibitem[{{Bundy} {et~al.}(2010){Bundy}, {Scarlata}, {Carollo}, {Ellis},
  {Drory}, {Hopkins}, {Salvato}, {Leauthaud}, {Koekemoer}, {Murray}, {Ilbert},
  {Oesch}, {Ma}, {Capak}, {Pozzetti}, \& {Scoville}}]{bundy10}
{Bundy}, K., {Scarlata}, C., {Carollo}, C.~M., {et~al.} 2010, \apj, 719, 1969

\bibitem[{{Catinella} {et~al.}(2010){Catinella}, {Schiminovich}, {Kauffmann},
  {Fabello}, {Wang}, {Hummels}, {Lemonias}, {Moran}, {Wu}, {Giovanelli},
  {Haynes}, {Heckman}, {Basu-Zych}, {Blanton}, {Brinchmann}, {Budav{\'a}ri},
  {Gon{\c c}alves}, {Johnson}, {Kennicutt}, {Madore}, {Martin}, {Rich},
  {Tacconi}, {Thilker}, {Wild}, \& {Wyder}}]{catinella10}
{Catinella}, B., {Schiminovich}, D., {Kauffmann}, G., {et~al.} 2010, \mnras,
  403, 683

\bibitem[{{Chabrier}(2003)}]{chabrier}
{Chabrier}, G. 2003, \pasp, 115, 763

\bibitem[{{Chilingarian} \& {Zolotukhin}(2012)}]{chilingarian12}
{Chilingarian}, I.~V. \& {Zolotukhin}, I.~Y. 2012, \mnras, 419, 1727

\bibitem[{{Cortese} \& {Hughes}(2009)}]{cortese09}
{Cortese}, L. \& {Hughes}, T.~M. 2009, \mnras, 400, 1225

\bibitem[{{Cowie} \& {Barger}(2008)}]{cowie08}
{Cowie}, L.~L. \& {Barger}, A.~J. 2008, \apj, 686, 72

\bibitem[{{Crowl} \& {Kenney}(2008)}]{crowl2008}
{Crowl}, H.~H. \& {Kenney}, J.~D.~P. 2008, \aj, 136, 1623

\bibitem[{{Franzetti} {et~al.}(2007){Franzetti}, {Scodeggio}, {Garilli},
  {Vergani}, {Maccagni}, {Guzzo}, {Tresse}, {Ilbert}, {Lamareille}, {Contini},
  {Le F{\`e}vre}, {Zamorani}, {Brinchmann}, {Charlot}, {Bottini}, {Le Brun},
  {Picat}, {Scaramella}, {Vettolani}, {Zanichelli}, {Adami}, {Arnouts},
  {Bardelli}, {Bolzonella}, {Cappi}, {Ciliegi}, {Foucaud}, {Gavignaud},
  {Iovino}, {McCracken}, {Marano}, {Marinoni}, {Mazure}, {Meneux}, {Merighi},
  {Paltani}, {Pell{\`o}}, {Pollo}, {Pozzetti}, {Radovich}, {Zucca}, {Cucciati},
  \& {Walcher}}]{franzetti07}
{Franzetti}, P., {Scodeggio}, M., {Garilli}, B., {et~al.} 2007, \aap, 465, 711

\bibitem[{{Gallazzi} {et~al.}(2009){Gallazzi}, {Bell}, {Wolf}, {Gray},
  {Papovich}, {Barden}, {Peng}, {Meisenheimer}, {Heymans}, {van Kampen},
  {Gilmour}, {Balogh}, {McIntosh}, {Bacon}, {Barazza}, {B{\"o}hm}, {Caldwell},
  {H{\"a}u{\ss}ler}, {Jahnke}, {Jogee}, {Lane}, {Robaina}, {Sanchez}, {Taylor},
  {Wisotzki}, \& {Zheng}}]{gallazzi09}
{Gallazzi}, A., {Bell}, E.~F., {Wolf}, C., {et~al.} 2009, \apj, 690, 1883

\bibitem[{{Gavazzi} {et~al.}(2002){Gavazzi}, {Bonfanti}, {Sanvito}, {Boselli},
  \& {Scodeggio}}]{gav02}
{Gavazzi}, G., {Bonfanti}, C., {Sanvito}, G., {Boselli}, A., \& {Scodeggio}, M.
  2002, \apj, 576, 135

\bibitem[{{Hao} {et~al.}(2011){Hao}, {Kennicutt}, {Johnson}, {Calzetti},
  {Dale}, \& {Moustakas}}]{hao11}
{Hao}, C.-N., {Kennicutt}, R.~C., {Johnson}, B.~D., {et~al.} 2011, \apj, 741,
  124

\bibitem[{{Heavens} {et~al.}(2004){Heavens}, {Panter}, {Jimenez}, \&
  {Dunlop}}]{heavens}
{Heavens}, A., {Panter}, B., {Jimenez}, R., \& {Dunlop}, J. 2004, \nat, 428,
  625

\bibitem[{{Kannappan} {et~al.}(2009){Kannappan}, {Guie}, \&
  {Baker}}]{kannappan09}
{Kannappan}, S.~J., {Guie}, J.~M., \& {Baker}, A.~J. 2009, \aj, 138, 579

\bibitem[{{Lintott} {et~al.}(2008){Lintott}, {Schawinski}, {Slosar}, {Land},
  {Bamford}, {Thomas}, {Raddick}, {Nichol}, {Szalay}, {Andreescu}, {Murray}, \&
  {Vandenberg}}]{lintott08}
{Lintott}, C.~J., {Schawinski}, K., {Slosar}, A., {et~al.} 2008, \mnras, 389,
  1179

\bibitem[{{Martin} {et~al.}(2005){Martin}, {Fanson}, {Schiminovich},
  {Morrissey}, {Friedman}, {Barlow}, {Conrow}, {Grange}, {Jelinsky},
  {Milliard}, {Siegmund}, {Bianchi}, {Byun}, {Donas}, {Forster}, {Heckman},
  {Lee}, {Madore}, {Malina}, {Neff}, {Rich}, {Small}, {Surber}, {Szalay},
  {Welsh}, \& {Wyder}}]{martin05}
{Martin}, D.~C., {Fanson}, J., {Schiminovich}, D., {et~al.} 2005, \apjl, 619,
  L1

\bibitem[{{Masters} {et~al.}(2010){Masters}, {Mosleh}, {Romer}, {Nichol},
  {Bamford}, {Schawinski}, {Lintott}, {Andreescu}, {Campbell}, {Crowcroft},
  {Doyle}, {Edmondson}, {Murray}, {Raddick}, {Slosar}, {Szalay}, \&
  {Vandenberg}}]{masters10}
{Masters}, K.~L., {Mosleh}, M., {Romer}, A.~K., {et~al.} 2010, \mnras, 405, 783

\bibitem[{{Moran} {et~al.}(2006){Moran}, {Ellis}, {Treu}, {Salim}, {Rich},
  {Smith}, \& {Kneib}}]{moran06}
{Moran}, S.~M., {Ellis}, R.~S., {Treu}, T., {et~al.} 2006, \apjl, 641, L97

\bibitem[{{Rowlands} {et~al.}(2012){Rowlands}, {Dunne}, {Maddox}, {Bourne},
  {Gomez}, {Kaviraj}, {Bamford}, {Brough}, {Charlot}, {da Cunha}, {Driver},
  {Eales}, {Hopkins}, {Kelvin}, {Nichol}, {Sansom}, {Sharp}, {Smith}, {Temi},
  {van der Werf}, {Baes}, {Cava}, {Cooray}, {Croom}, {Dariush}, {de Zotti},
  {Dye}, {Fritz}, {Hopwood}, {Ibar}, {Ivison}, {Liske}, {Loveday}, {Madore},
  {Norberg}, {Popescu}, {Rigby}, {Robotham}, {Rodighiero}, {Seibert}, \&
  {Tuffs}}]{rowlands12}
{Rowlands}, K., {Dunne}, L., {Maddox}, S., {et~al.} 2012, \mnras, 419, 2545

\bibitem[{{Salim} {et~al.}(2007){Salim}, {Rich}, {Charlot}, {Brinchmann},
  {Johnson}, {Schiminovich}, {Seibert}, {Mallery}, {Heckman}, {Forster},
  {Friedman}, {Martin}, {Morrissey}, {Neff}, {Small}, {Wyder}, {Bianchi},
  {Donas}, {Lee}, {Madore}, {Milliard}, {Szalay}, {Welsh}, \& {Yi}}]{salim07}
{Salim}, S., {Rich}, R.~M., {Charlot}, S., {et~al.} 2007, \apjs, 173, 267

\bibitem[{{Schawinski} {et~al.}(2009){Schawinski}, {Lintott}, {Thomas},
  {Sarzi}, {Andreescu}, {Bamford}, {Kaviraj}, {Khochfar}, {Land}, {Murray},
  {Nichol}, {Raddick}, {Slosar}, {Szalay}, {Vandenberg}, \&
  {Yi}}]{schawinski09}
{Schawinski}, K., {Lintott}, C., {Thomas}, D., {et~al.} 2009, \mnras, 396, 818

\bibitem[{{Scodeggio} {et~al.}(2002){Scodeggio}, {Gavazzi}, {Franzetti},
  {Boselli}, {Zibetti}, \& {Pierini}}]{scodeggio02}
{Scodeggio}, M., {Gavazzi}, G., {Franzetti}, P., {et~al.} 2002, \aap, 384, 812

\bibitem[{{Thomas} {et~al.}(2010){Thomas}, {Maraston}, {Schawinski}, {Sarzi},
  \& {Silk}}]{thomas10}
{Thomas}, D., {Maraston}, C., {Schawinski}, K., {Sarzi}, M., \& {Silk}, J.
  2010, \mnras, 404, 1775

\bibitem[{{van den Bergh}(1976)}]{vandberg76}
{van den Bergh}, S. 1976, \apj, 206, 883

\bibitem[{{Visvanathan}(1981)}]{visvanathan81}
{Visvanathan}, N. 1981, \aap, 100, L20

\bibitem[{{Visvanathan} \& {Sandage}(1977)}]{visvanathan77}
{Visvanathan}, N. \& {Sandage}, A. 1977, \apj, 216, 214

\bibitem[{{Wolf} {et~al.}(2009){Wolf}, {Arag{\'o}n-Salamanca}, {Balogh},
  {Barden}, {Bell}, {Gray}, {Peng}, {Bacon}, {Barazza}, {B{\"o}hm}, {Caldwell},
  {Gallazzi}, {H{\"a}u{\ss}ler}, {Heymans}, {Jahnke}, {Jogee}, {van Kampen},
  {Lane}, {McIntosh}, {Meisenheimer}, {Papovich}, {S{\'a}nchez}, {Taylor},
  {Wisotzki}, \& {Zheng}}]{wolf09}
{Wolf}, C., {Arag{\'o}n-Salamanca}, A., {Balogh}, M., {et~al.} 2009, \mnras,
  393, 1302

\bibitem[{{Wright} {et~al.}(2010){Wright}, {Eisenhardt}, {Mainzer}, {Ressler},
  {Cutri}, {Jarrett}, {Kirkpatrick}, {Padgett}, {McMillan}, {Skrutskie},
  {Stanford}, {Cohen}, {Walker}, {Mather}, {Leisawitz}, {Gautier}, {McLean},
  {Benford}, {Lonsdale}, {Blain}, {Mendez}, {Irace}, {Duval}, {Liu}, {Royer},
  {Heinrichsen}, {Howard}, {Shannon}, {Kendall}, {Walsh}, {Larsen}, {Cardon},
  {Schick}, {Schwalm}, {Abid}, {Fabinsky}, {Naes}, \& {Tsai}}]{wright10}
{Wright}, E.~L., {Eisenhardt}, P.~R.~M., {Mainzer}, A.~K., {et~al.} 2010, \aj,
  140, 1868

\bibitem[{{Wyder} {et~al.}(2007){Wyder}, {Martin}, {Schiminovich}, {Seibert},
  {Budav{\'a}ri}, {Treyer}, {Barlow}, {Forster}, {Friedman}, {Morrissey},
  {Neff}, {Small}, {Bianchi}, {Donas}, {Heckman}, {Lee}, {Madore}, {Milliard},
  {Rich}, {Szalay}, {Welsh}, \& {Yi}}]{wyder07}
{Wyder}, T.~K., {Martin}, D.~C., {Schiminovich}, D., {et~al.} 2007, \apjs, 173,
  293

\bibitem[{{Yi} {et~al.}(2005){Yi}, {Yoon}, {Kaviraj}, {Deharveng}, {Rich},
  {Salim}, {Boselli}, {Lee}, {Ree}, {Sohn}, {Rey}, {Lee}, {Rhee}, {Bianchi},
  {Byun}, {Donas}, {Friedman}, {Heckman}, {Jelinsky}, {Madore}, {Malina},
  {Martin}, {Milliard}, {Morrissey}, {Neff}, {Schiminovich}, {Siegmund},
  {Small}, {Szalay}, {Jee}, {Kim}, {Barlow}, {Forster}, {Welsh}, \&
  {Wyder}}]{yi05}
{Yi}, S.~K., {Yoon}, S.-J., {Kaviraj}, S., {et~al.} 2005, \apjl, 619, L111

\end{thebibliography}

\end{document}